\documentclass[12pt]{iopart}
\usepackage{graphicx}
\usepackage{natbib}
\bibpunct{(} {)} {;} {} {} {,}

\begin{document}

\title[Mammographic image restoration using maximum entropy
deconvolution]{Mammographic image restoration using maximum
entropy deconvolution}

\author{A Jannetta and J C Jackson \S}
\address{\S\ School of Informatics, Northumbria University,
Newcastle-upon-Tyne, NE1 8ST, UK}

\author{C J Kotre, I P Birch, K J Robson and R
Padgett \dag}

\address{\dag\ Regional Medical Physics Group, Newcastle General Hospital, Newcastle-upon-Tyne, NE4 6BE, UK}

\ead{john.jackson@unn.ac.uk}

\begin{abstract}
An image restoration approach based on a Bayesian maximum entropy
method (MEM) has been applied to a radiological image deconvolution
problem, that of reduction of geometric blurring in magnification
mammography.  The aim of the work is to demonstrate an improvement
in image spatial resolution in realistic noisy
radiological images with no associated penalty in terms of
reduction in the signal-to-noise ratio perceived by the observer.
Images of the TORMAM mammographic image quality phantom were
recorded using the standard magnification settings of 1.8
magnification/fine focus and also at 1.8 magnification/broad
focus and 3.0 magnification/fine focus; the latter two arrangements
would normally give rise to unacceptable geometric blurring.
Measured point-spread functions were used in conjunction with the
MEM image processing to de-blur these images. The results are
presented as comparative images of phantom test features and as
observer scores for the raw and processed images. Visualization of
high resolution features and the total image scores for the test
phantom were improved by the application of the MEM processing. It
is argued that this successful demonstration of image de-blurring
in noisy radiological images offers the possibility of weakening
the link between focal spot size and geometric blurring in
radiology, thus opening up new approaches to system optimization.
\end{abstract}

\section {Introduction}
The design of radiological imaging equipment has developed to
reflect the best compromise between a number of contradictory
performance requirements. In particular the design of x-ray tubes
has evolved to best fit the requirements for the smallest possible
focal spot size combined with the maximum target heat capacity. In
this case the performance compromise is between geometrical
blurring introduced by a finite focal spot size and movement
blurring introduced by the long radiographic exposure time
required to operate a very small focal spot within its heat
loading limits. The most common resulting design, rotating target
with an acute target angle, represents an impressive engineering
solution to this problem.

An alternative potential solution to the problem of geometrical
blurring, applicable to modern digital receptors, is digital image
processing. Provided the characteristics of the image blurring
function are known, an image can, in theory, be restored to that
which would have been produced by a perfect point focal spot.
However, in practice the degree to which radiological images can
be restored to remove geometric blurring, using conventional
Fourier deconvolution strategies, is severely limited by noise.
The relatively poor signal-to-noise ratio of radiological images
is a direct result of the requirement to operate within the ALARP
principle, so that patient radiation doses are `kept as low as
reasonably practicable consistent with the intended purpose'
\citep{IRMER2000}. Radiological images are therefore expected to
be inherently noisy, with quantum noise being the dominant noise
source. The noise power extends to the highest spatial frequencies
recorded by the receptor, and it is the modulation at these higher
spatial frequencies which is amplified using Fourier-based
techniques, resulting in an image dominated by high frequency
noise. The Wiener filter \citep{WIENER,HELSTROM} is a popular
deconvolution tool which attempts to address the shortcomings of
direct Fourier inversion techniques.  Its application to
non-mammographic images has been demonstrated by \cite{DOUGH}.
However, it is our experience that these techniques are of
marginal value in radiology.

In this work, an alternative image restoration approach based on a
Bayesian maximum entropy method is applied to a radiological image
deconvolution problem: reduction of geometric blurring in mammography.
The aim of the work is to demonstrate an improvement in image
spatial resolution with no associated penalty in terms
of reduction in the signal-to-noise ratio perceived by
the observer.  The images used here were digitized \textit{ab initio};
in principle digitized film could be used, but this would almost
certainly be an unacceptable inconvenience if digital processing
of this nature were to become part of clinical practice.  Indeed
it is the advent of Computed and Direct Radiography which we believe
make this work particularly timely.

\section {Medical image restoration} \label{MIR}

There is a large body of literature describing methods which aim
to improve the quality of image information content through
restoration techniques.  Medical images are usually measurements
of photon flux, and the data is usually noisy and often
incomplete; statistical methods have yielded some robust methods
of estimating the `true' image distribution in these
circumstances.  The Expectation Maximization (EM) algorithm
\citep{DEMPSTER} has been widely applied to finding solutions
which maximize a likelihood function for quantum noise limited
data; in the medical context this method has been used by
\citet{HUDSON}, \citet{DEPIERRO} and \citet{KINAHAN} for
tomographic imaging.  In the astronomical community EM is known as
Richardson-Lucy deconvolution \citep{RICHARDSON,LUCY,SHEPP}.
\Citet{SNYDER1} describe a method which maximizes the mean value
of the log-likelihood for quantum noise limited data. It has been
shown that this is equivalent to minimizing Csisz\`{a}r's
I-divergence \citep{CSISZAR}, a quantity equal to the negative of
the entropy expression, $-S$, given in equation (\ref{Entropy}).
The usefulness of Bayesian restoration stems from the fact that it
allows the incorporation of sophisticated \textit{a priori}
knowledge into the formulation of the restoration method, while
quite naturally enforcing desirable properties such as positivity
in the restoration.  It has been argued by \citet{SKILLING2} that
in the absence of further prior information entropy is the only
consistent prior for positive, additive images. \Citet{OSULLIVAN}
gives a summary of these methods in terms of information theoretic
image formation models.

\subsection {MEM theory}

The links between statistical mechanics and information theory
were established by \cite{JAYNES4,JAYNES5,JAYNES2}.
Image restoration using MEM was first described in a landmark paper by
\cite{FRIEDEN}. The driving force behind practical
implementations of the method came from radio astronomy
and the need to improve radio maps of the sky \citep{GULLDAN, CORNEVANS}.

MEM is a deconvolution technique derived from the forward map (1)
for the imaging system, which relates postulated hidden data
$x_{ij}$  to the observed data $d_{ij}$.

\begin{figure}[here]
\begin{center}
\includegraphics[scale=0.5]{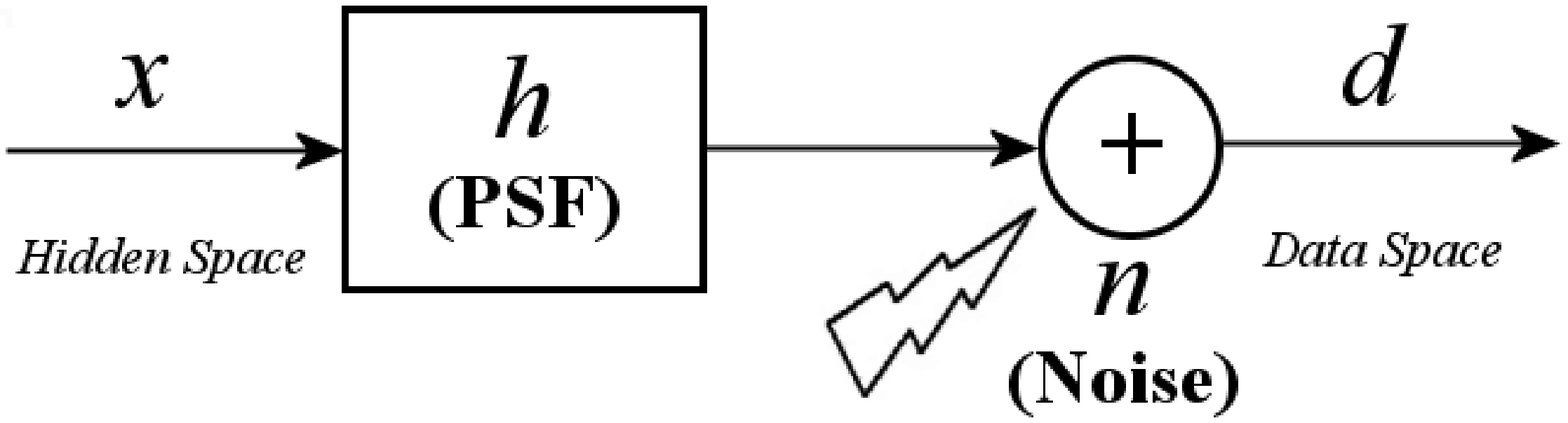}
\end{center}
\end{figure}

\begin{equation}\label{forwardmapfig}
    d_{ij}=x_{ij}\ast h + n_{ij}
\end{equation}

Equation (\ref{forwardmapfig}) is the forward map for the imaging
system.  The PSF, denoted by $h$, characterizes the geometric
blurring of the imaging system, which acts on the hidden data. The
hidden image is then further corrupted by additive noise to
produce the observed image $d_{ij}$.

The goal of image restoration is to obtain a solution $\hat{x}$ ,
which approximates to the hidden image $x$, as closely as the data
and noise allow.  We outline briefly the MEM approach to the
problem here, but for a more complete review see \cite{JAYNES3}
and \cite{GULLSKILL}. The notation in the following analysis has
been simplified in relation to equation (\ref{forwardmapfig}), in
that the data etc. are represented as vectors rather than 2D
arrays.

A trial restoration $\hat{x}$ is obtained and used as an initial
guess for the hidden image $x$.  The trial restoration is blurred,
via equation (\ref{forwardmapfig}), to generate mock data $\hat{d}$.
The $\chi^{2}$ goodness of fit statistic is used to measure the degree
of misfit between the observed data and mock data:

\begin{equation}\label{chi2}
    \chi^{2} = \frac{\sum _{i} (d_{i} - \hat{d_{i}})^{2}}{\sigma^{2}}
\end{equation}

\noindent where $\sigma^{2}$ is the variance in the noise, here
taken to be constant.  It might be thought that a good approach
would be to minimize this degree of misfit by choosing a suitable
$\hat{x}$, but this process is equivalent to the straightforward
matrix inversion:

\begin{equation}\label{inversion}
    \hat{x} = A^{-1}d
\end{equation}

\noindent where $A$ is the matrix representation of the linear
forward map equation (\ref{forwardmapfig}).  The problem here is
that typically the matrix $A$ is ill-conditioned, i.e. almost
singular, so for a given $d$ there are many vectors $\hat{x}$
which almost satisfy equation (\ref{inversion}), not necessarily
close together.  Thus if $d$ contains even a small amount of
noise, the resulting $\hat{x}$ can be far from the true image $x$;
in other words the reconstruction is then dominated by noise
rather than data, often referred to as over-restoration. In the
presence of noise we would not in any case expect $\chi^{2}$ to be
minimized, but rather to be reduced to the appropriate value
$\chi^{2}=N$, where $N$ is the number of pixels in the image. Even
when $A$ is not ill-conditioned there are many $\hat{x}$ which
satisfy this criterion.  MEM is an example of regularized
deconvolution and relies on a scheme of iterated forward maps
rather than attempting to find a direct solution of the inverse
problem.

MEM treats the restoration process as a statistical inference
problem based on Bayes' theorem and the aim is to obtain the most
probable image $\hat{x}$ given the data:

\begin{equation}\label{Bayes}
    P(\hat{x}\mid d)\propto P(d\mid \hat{x}) \times P(\hat{x})
\end{equation}

\noindent The likelihood $P(d\mid \hat{x})$ is determined from our
knowledge of the forward map equation (\ref{forwardmapfig}); the
image noise is mainly quantum (photon) noise, which is modelled as
a Gaussian process, so the likelihood term is quantified by the
$\chi^{2}$ distribution:

\begin{equation}\label{likelihood}
    P(d\mid \hat{x}) \propto \exp (- \chi^{2}/2)
\end{equation}

\noindent The prior $P(\hat{x})$ is the probability that would be
assigned by to a particular reconstruction $\hat{x}$ prior to the
introduction of observational constraints.  It is assumed that
each luminance quantum has an equal \textit{a priori} chance of
falling into any pixel, in which case it is easy to show that

\begin{equation}\label{prior}
    P(\hat{x}) \propto \exp (S)
\end{equation}

\noindent where $S$ is the configurational entropy of the hidden
image.  For positive additive distributions the entropy is defined
as \citep{SKILLING}:

\begin{equation}\label{Entropy}
    S(\hat{x})=\sum_{i} [\hat{x_{i}} - m - \hat{x_{i}}\ln (\hat{x_{i}}/m)]
\end{equation}

\noindent Here $m$ is a (constant) default level; $S$ is maximized
when $\hat{x_{i}}=m$, giving a flat featureless reconstruction. If
each luminance quantum has a different chance of falling into any
pixel, the default levels $m_{i}$  for pixel $i$ can be chosen
accordingly, and S is maximized when $\hat{x_{i}}=m_{i}$.  A
non-flat default level would be appropriate for example when the
X-ray illumination is not uniform or when other knowledge of the
restoration is known \textit{a priori}.  The strict definition of posterior
probability is therefore:

\begin{equation}\label{posterior1}
    P(\hat{x}\mid d)\propto \exp (S - \chi^{2}/2)
\end{equation}

Values of $\hat{x_{i}}$ which maximize this probability should be
determined.  This cannot be achieved analytically, and a numerical
scheme must be employed.  In practice a pragmatic modification of
equation (\ref{posterior1}) is used:

\begin{equation}\label{posterior2}
    P(\hat{x}\mid d)\propto \exp (\alpha S - \chi^{2})= \exp (Q)
\end{equation}

\noindent where $Q=\alpha S - \chi^{2}$, and $\alpha$ is a
multiplier, as yet undetermined. $\chi^{2}$ is a measure of the
misfit between the actual observations and those corresponding to
the trial values $\hat{x_{i}}$; $S$ is a measure of the structure
within the trial image. If $\alpha$ is too small then too much
weight will be given to the data, thus amplifying the effects of
noise; if $\alpha$ is too large then too much weight will be given
to the entropy and each $\hat{x_{i}}$ will be dragged towards its
default value $m_{i}$, thus losing real features. The procedure is
to maximize $Q$ with $\alpha$ initially fixed; a particular value
of $\chi^{2}$ will correspond to this maximum.  The aim is that
this value should coincide with the expectation $\chi^{2}=N$; the
value of $\alpha$ which achieves this aim is found by a process of
iteration.

Initial feasibility tests were undertaken using the MATLAB
Optimization Toolbox (The Mathworks Ltd, Cambridge, UK),
particularly the constrained nonlinear minimization function {\it
fmincon} in its LargeScale configuration, with a positivity
constraint on each $\hat{x_{i}}$. This function employs a
preconditioned conjugate gradient method
\citep{COLEMAN2,COLEMAN1}. Our procedure has two loops. The inner
one, in which $\alpha$ is fixed, iterates towards the minimum
value of $-Q$ for that $\alpha$, and generates a corresponding
value $\chi^2(\alpha)$. The second loop iterates over $\alpha$ to
minimize $\chi^2(\alpha)-N$, and is terminated when
$\chi^2(\alpha)$ falls within the narrow statistical range $N \pm
(2N)^{1/2}$ \citep{JACKSON}. Typically each loop requires 20 to 30
iterations, and the final value of $\alpha$ is of order 20. After
our successful initial implementation commercially available C++
software was purchased (MEMSYS5, Maximum Entropy Data Consultants,
Bury St Edmunds, UK).  This employs essentially the same scheme,
but is highly optimized, as described in \citet{SKILLINGBRYAN} and
\citet{GULLSKILL2}. The MEMSYS5 software can handle larger images
than our initial implementation, with much shorter processing
times.

\section {Experimental Method}

\begin{figure}
\begin{center}
\includegraphics[scale=0.8]{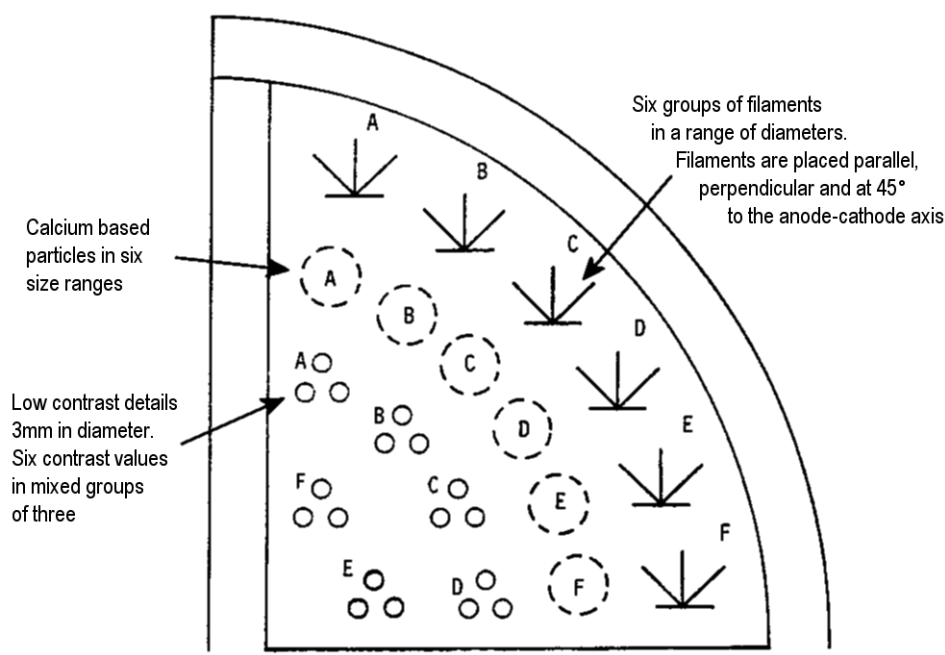}
\caption{Layout of the quantitative side of the Leeds TORMAM test object}
\label{figureTORMAM}
\end{center}
\end{figure}

Image quality comparisons were carried out using the Leeds TORMAM
test object (figure \ref{figureTORMAM}) at various settings of
geometric magnification. This phantom contains three groups of
test features; fibres, simulated microcalcification clusters and
low contrast plastic disks, plus an area designed to give an
anthropomorphic impression of a breast parenchymal pattern with
overlying microcalcification clusters \citep{COWEN}. This latter
area of the phantom was not used in the present work. The phantom
was imaged in various configurations on a laboratory-based Siemens
Mammomat 3 mammography unit, with measured focal spot sizes of
0.7$\times$0.3mm broad focus  and 0.2$\times$0.2mm fine focus. In
all cases the phantom was imaged on top of a 3.5cm thick stack of
Perspex to provide realistic scatter and attenuation as
recommended in the instructions for the phantom. The radiographic
factors used were 28kVp, 40mAs. These factors were chosen to be
representative of the values used in routine mammographic quality
assurance tests. The level of quantum noise in the test images was
therefore realistic. The digital image receptor was part of
a Philips ACR-3 Computed Radiography system (100 $\mu$m pixel size),
comprising $1770 \times 2370$ pixels with overall dimensions $18 \times 24$ cm.

\begin{figure}[here]
\begin{center}
\includegraphics[scale=0.85]{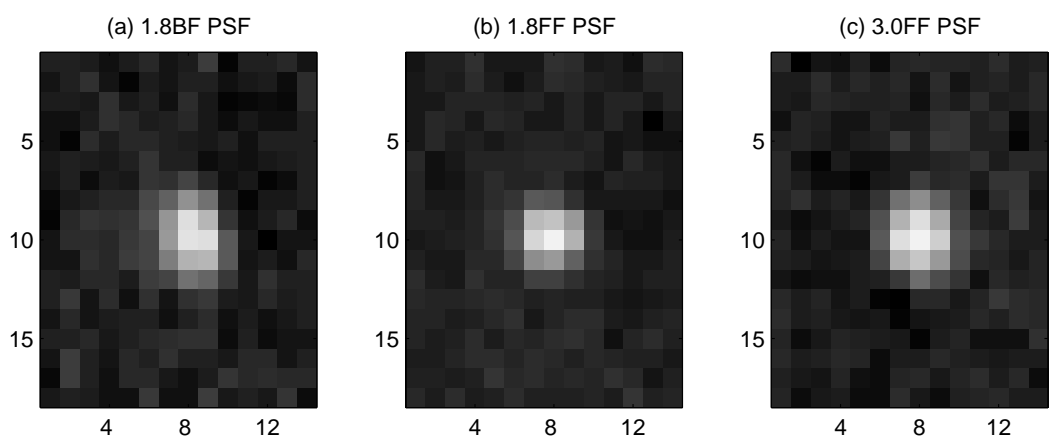}
\caption{Raw images of the focal spot obtained under the
configurations (a) 1.8BF, (b) 1.8FF and (c) 3.0FF.}
\label{figurePSF}
\end{center}
\end{figure}

In order to provide a sample of the point spread function (PSF) in
the plane of the phantom, a piece of brass foil with a pinhole was
included in the test images. In any practical application of this
technique, the PSF information would be obtained from calibration
images appropriately scaled for the position of the object.
Including the PSF with the image was convenient in this
demonstration as the scaling and interpolation step was avoided.
The measured overall PSF is a convolution of one due to geometry
and the the detector PSF, but is nevertheless the appropriate
one to use.  In fact the contribution from the detector PSF was small,
evidenced by the fact that the measured PSF scaled with focal spot size
and magnification in the appropriate manner (see figure \ref{figurePSF}).

Three imaging geometries were used to illustrate varying degrees
of focal spot geometrical blurring:

\begin{description}
    \item[Magnification 1.8, broad focus (1.8BF).] The plane of the phantom was 33.5cm from the tube focus and 27cm from the plane of the receptor. The image would be expected to be unacceptably degraded by geometric blurring.
    \item[Magnification 1.8, fine focus (1.8FF).] The phantom was in the same position as above, but imaged using the fine focal spot. This is the conventional magnification view provided on this mammography unit and used clinically.
    \item[Magnification 3.0, fine focus (3.0FF).] The plane of the phantom was 20.5cm from tube focus and 40cm from the plane of the receptor.  This high magnification factor is not supported on standard mammography units because even using fine focus, the image would be unacceptably blurred.
\end{description}

Images of the focal spots obtained in each configuration are shown
in figure \ref{figurePSF}.

The original and processed sets of images were viewed and scored
by two independent observers both experienced in the use of
mammographic image quality test phantoms. The test images were
graded using the 3, 2, 1, 0 scoring system recommended in the
TORMAM phantom instructions and adopted in surveys of mammographic
image quality in the UK Breast Screening Programme
\citep{NHSBSP2003}.

\section {Image processing procedure}

\suppressfloats
\begin{figure}[top]
\begin{center}
\includegraphics[scale=0.75]{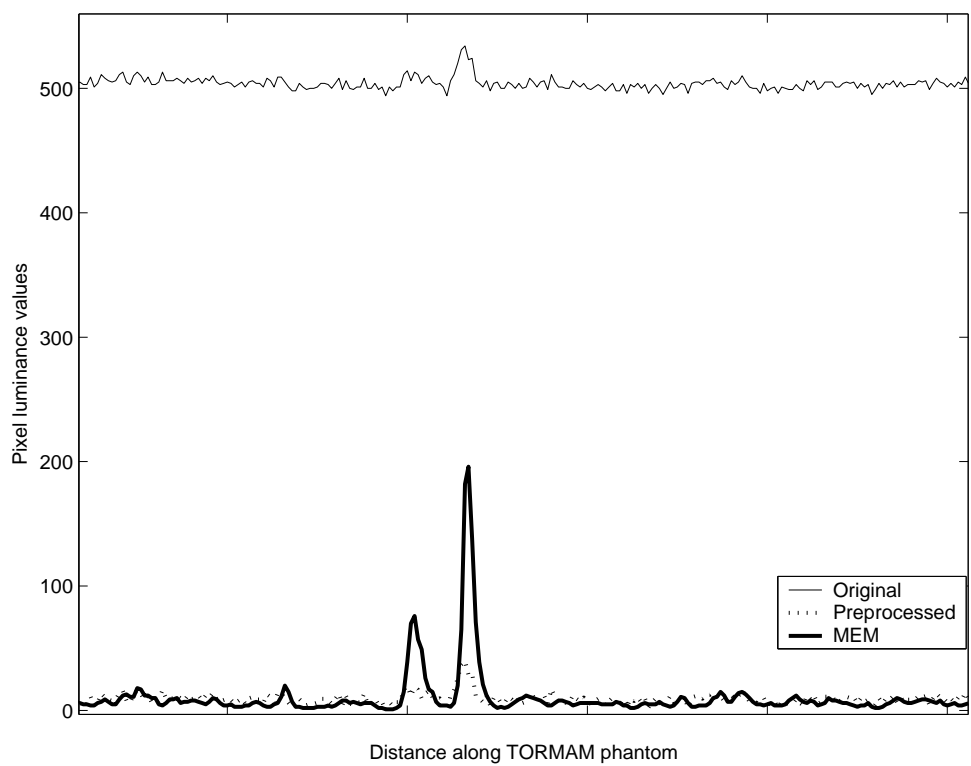}
\caption{1.8BF. Pixel intensity profiles through particles in
group A. The preprocessing step reduces and flattens the varying
background of the image but leaves the structure and noise intact.
MEM processing smoothes the noise and increases the
signal-to-noise ratio of features deemed statistically to be
caused by real objects. }
\label{figurePROFILE}
\end{center}
\end{figure}

The original images were in Siemens SPI file format which were
12-bit grayscale images holding 10-bit image data.  A conversion
to DICOM format was necessary to allow MATLAB to read the image
data, but the conversion process retained all the luminance
information held in the original files.

MATLAB R12 was used to perform some preprocessing so that the subsequent
MEM processing could be applied.  These steps were:

\begin{itemize}
    \item The original images have a varying background brightness caused by
non-uniform  intensity of the X-ray beam (which is by design in mammographic systems)
and by variations in sensitivity over each digital plate (several such were used).
To make the images more amenable to MEM processing it was desirable to reduce this
background \citep{DONOHO3}.  A Gaussian low pass filter was constructed and applied
to the original images in the Fourier domain to obtain a background map.  This map was subtracted from the original leaving a flattened image, which retained the phantom's high frequency features (see figure \ref{figurePROFILE}).  It should be noted that the Philips ACR-3 system is not a linear one, as each pixel value is
proportional to the logarithm of the number of photons striking the detector;
however, as the  raw signal to mean background ratio is small (see figure \ref{figurePROFILE}), typically less than  0.1, this procedure effectively
linearizes the system, necessary for application of any of the methods
discussed in section \ref{MIR}.
    \item Removal of spurious, bright pixels not corresponding to any real feature in the TORMAM phantom.  These pixels would lead to artefacts in the restored image.  The removal
was done by observing the image histogram and setting high valued outliers to some mean background level.
    \item The PSF images were cropped from the original TORMAM images.  In each case the
noisy background of the PSF was filtered out (typically by discarding those pixels with
less than 10\% of the PSF peak value).  This had the effect of slightly narrowing each
PSF - thus leading to a conservative under-restoration.
\end{itemize}

MEMSYS5 was used to treat the images following the MATLAB
preprocessing.  The interface to the MEMSYS5 kernel accepts image
and PSF-image files as inputs (MATLAB image format was convenient)
and allows the setting of certain parameters related to the theory
described in section 2.  The default image model $m$ was defined
to be a flat image with low pixel luminance values: our belief
that, in the absence of data, very high photon counts were
recorded (i.e. no absorption due to intervening material). MEMSYS5
uses a slightly more sophisticated forward map that that presented
in equation (\ref{forwardmapfig}).

The processed images contained between 2.2 million and 3.9 million
pixels.  MEMSYS5 typically converged to a solution within 15
$\alpha$-iterations with a processing time of four to eight minutes,
for an image of the full test object shown in figure \ref{figureTORMAM}.
The smaller images to be presented in figures \ref{figureCROW1} to
\ref{figureCROW2} were cropped from such an image after processing,
rather than processed individually.  The processed output files
from MEMSYS5 were 8-bit PNG files.  All image processing was carried
out on a Pentium 4 2.4GHz machine with 512MB of RAM.

\section {Results}
\setcounter{table}{0}

\begin{figure}[here]
\begin{center}
\includegraphics[scale=1]{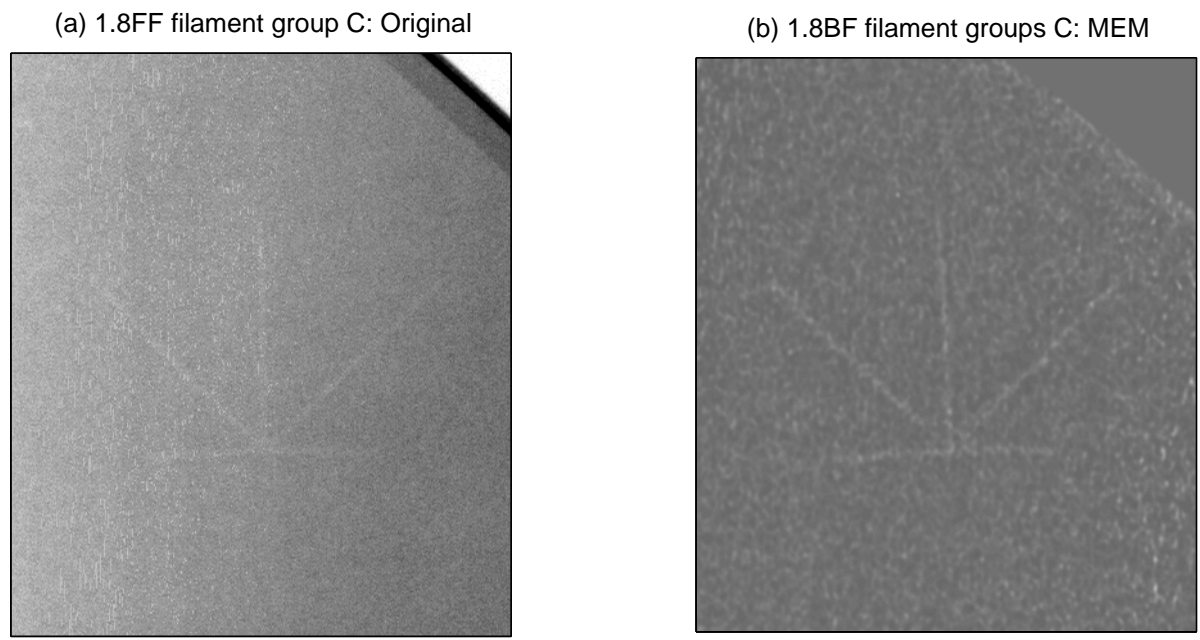}
\caption{Comparison of the filament group C imaged under 1.8FF
with the same feature imaged under 1.8BF and processed with MEM.}
\label{figureCROW1}
\end{center}
\end{figure}

\begin{figure}[here]
\begin{center}
\includegraphics[scale=1]{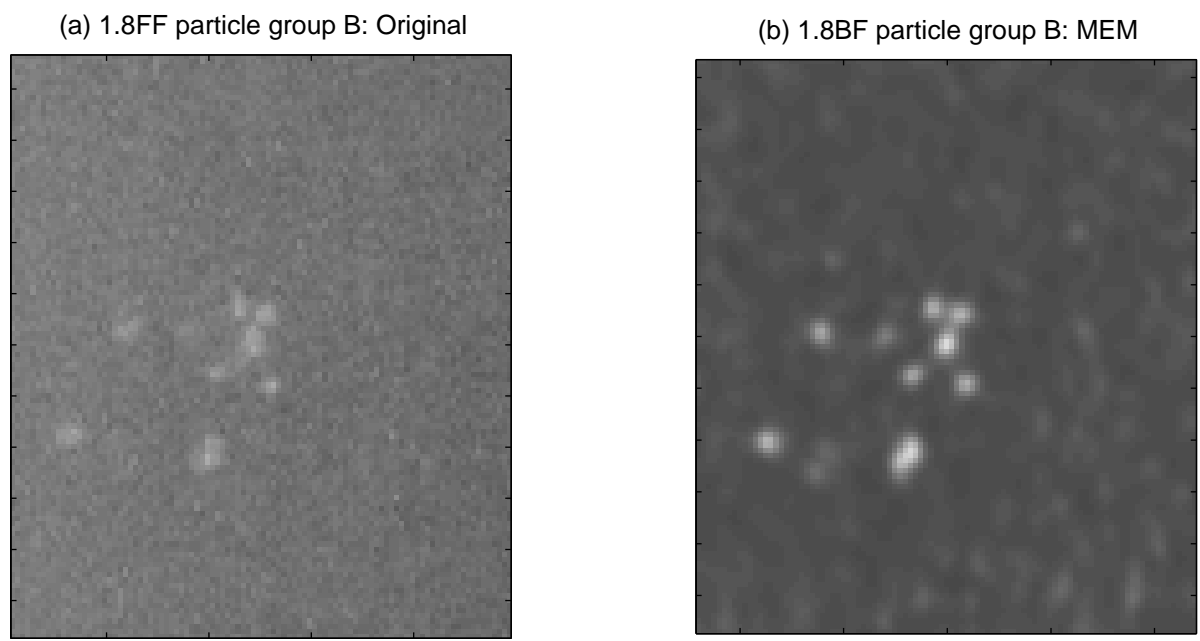}
\caption{Comparison of particle group B imaged under 1.8FF
with the same feature imaged under 1.8BF and processed with MEM.}
\label{figureCALC}
\end{center}
\end{figure}

Before proceeding to a systematic comparison of the original and
MEM processed images and related scores, we will say a few words
about our initial aspirations for this technique. As mentioned in
Section 1 these were to show improved spatial resolution without
reduction in signal-to-noise ratio perceived by the observer,
particularly a level of performance in which MEM processed 1.8BF
images are at least as good as unprocessed 1.8FF ones. Our belief
is that such a performance would be of clinical interest. Figures
\ref{figureCROW1} and \ref{figureCALC} compare appropriate images
of a filament group and simulated microcalcifications.  These
figures show that MEM reconstruction can give improvements in both
resolution {\it and} perceived signal-to-noise ratio, and that the
MEM processed 1.8BF images are marginally better in this instance
than the 1.8FF unprocessed ones in both respects. This promise
encouraged us to undertake a systematic evaluation, and to
consider magnifications greater than those normally used in
clinical practice. Three particle groups (B,D and E) were chosen
for this evaluation. The nominal particle size ranges for these
groups are 180-283 $\mu$m, 106-177 $\mu$m and 90-141 $\mu$m
respectively \citep{COWEN}.  In addition to the scoring system
mentioned earlier we also calculate some signal to noise ratios
(SNR) for these particular features.

\begin{figure}
\begin{center}
\includegraphics[scale=1]{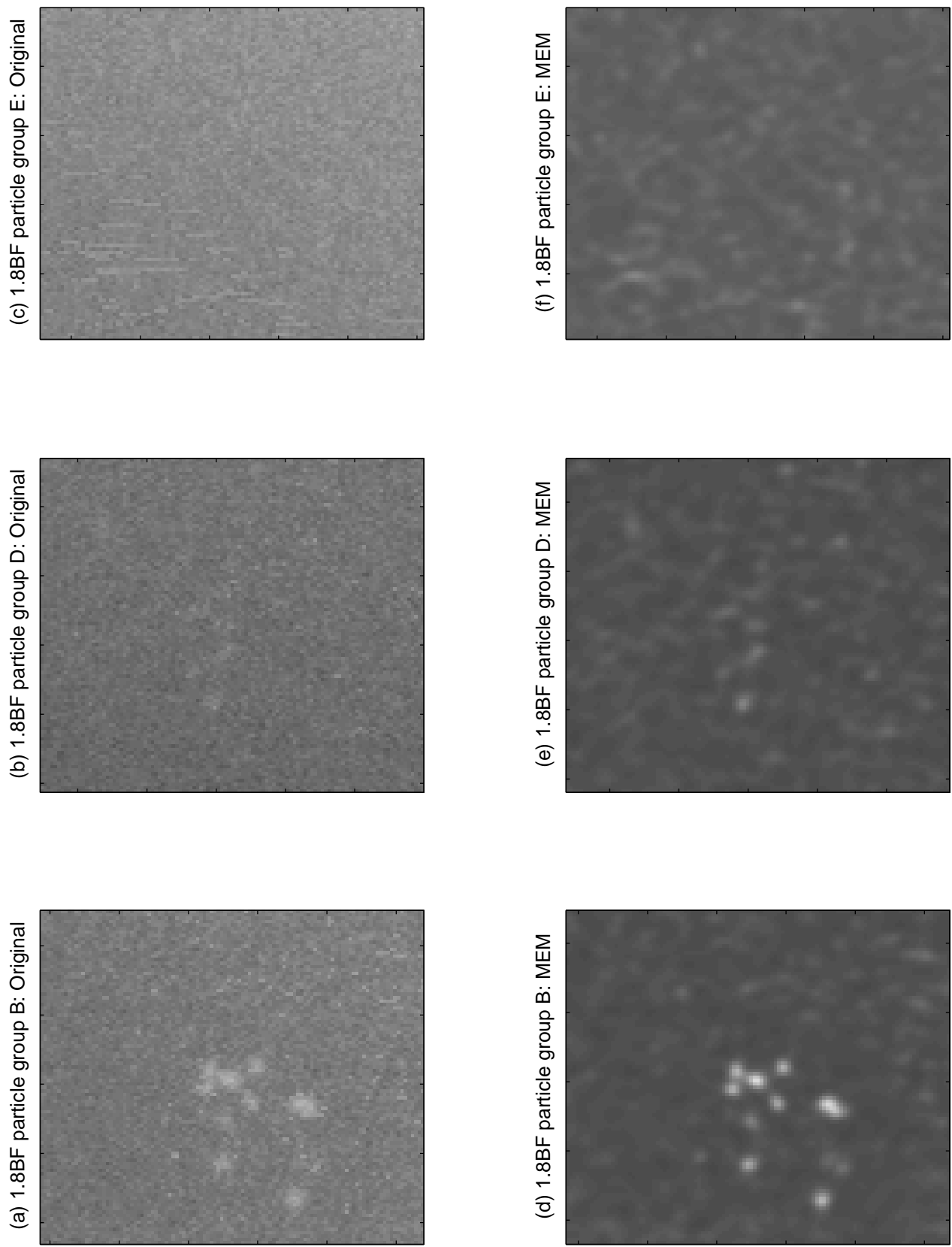}
\caption{1.8BF. Close-ups of particle groups B, D and E are shown
in (a), (b) and (c) respectively.  MEM restorations are shown
beneath in (d), (e) and (f); improvements in visibility and resolution
are apparent with the first two groups.  Group E is not detectable in
either the original or MEM processed images.}
\label{figure1.8BF}
\end{center}
\end{figure}

Figure \ref{figure1.8BF} shows images of the three particle
groups, cropped from the images taken under 1.8BF conditions.
The selected particle group is clearly visible in figures \ref{figure1.8BF}
(a) and (d), discernible in figure \ref{figure1.8BF} (e), just discernible
in figure \ref{figure1.8BF} (b) and not seen in figures \ref{figure1.8BF}
(c) and (f); however, we show the latter images to preserve the two by
three format, which eases comparison with later improvements.
As expected, unprocessed images obtained in this configuration are too
blurred to be clinically useful.  However, the MEM processed images show
significant improvements in resolution and signal-to-noise ratio,
particularly regarding the high frequency noise, which were quantified
using cuts similar to that shown in figure \ref{figurePROFILE}; the quoted
SNR is the difference between the largest signal and the mean background
within each group, divided by the standard deviation of the signal in a region close
to the group.  Figures for Group B are $\mathrm{{SNR}_{orig}}=9.0$ and $\mathrm{{SNR}_{MEM}}=50.3$.  Group D: $\mathrm{{SNR}_{orig}}=5.9$ and $\mathrm{{SNR}_{MEM}}=13.7$.  Particle group E is not detected at this setting.

\begin{figure}
\begin{center}
\includegraphics[scale=1]{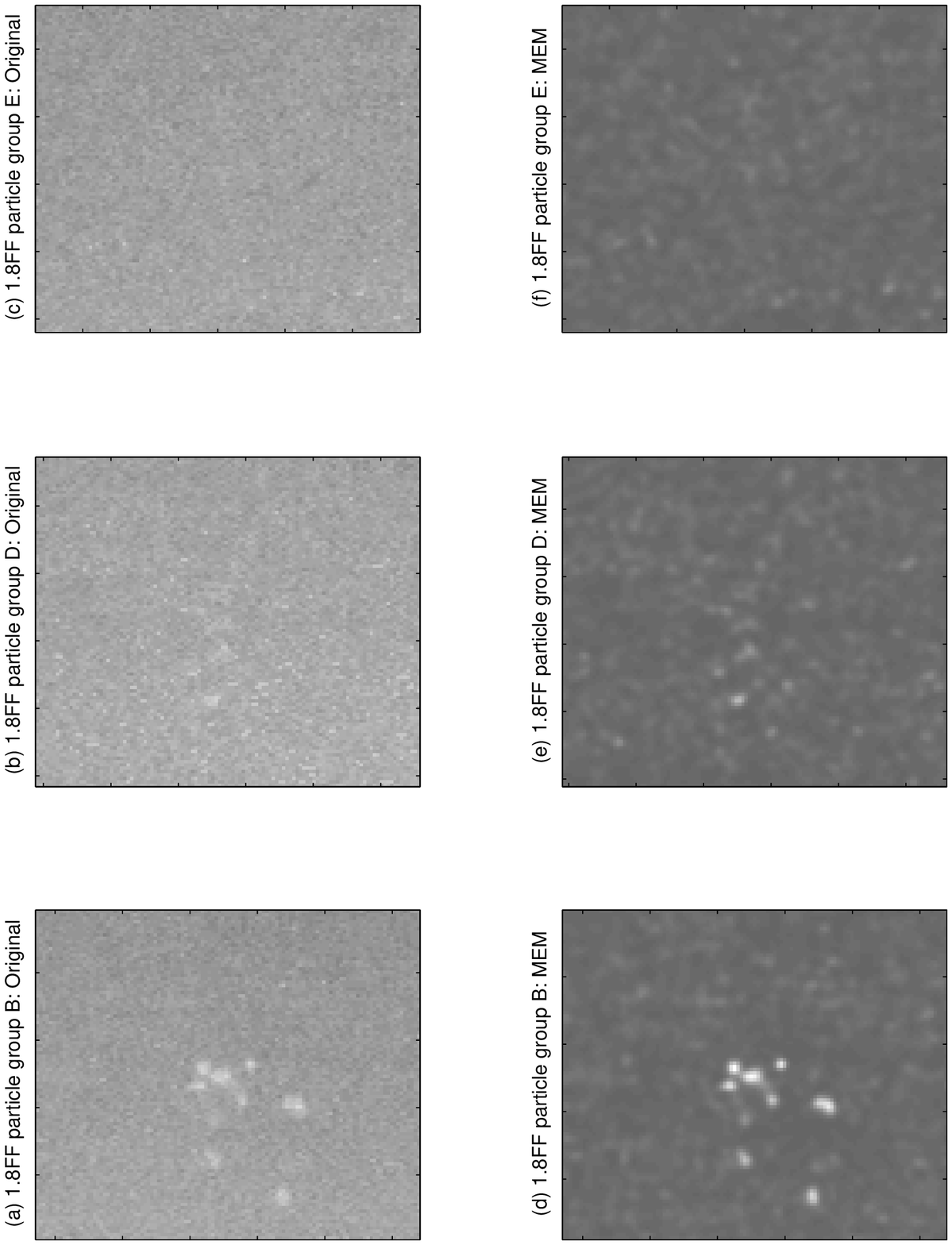}
\caption{1.8FF. Close-ups of particle groups B, D and E are shown
in (a), (b) and (c) respectively.  MEM restorations are shown
beneath in (d), (e) and (f); improvements in visibility and resolution
are apparent with the first two groups.  Group E is not detectable in
either the original or MEM processed images.}
\label{figure1.8FF}
\end{center}
\end{figure}

Figure \ref{figure1.8FF} shows cropped images of the same three
particle groups, obtained with the conventional 1.8FF
configuration.   Remarks regarding group visibility are as for
figure \ref{figure1.8BF}.  The pinhole PSF imaged under these conditions is
small and approximately Gaussian in shape, comprising just a few
pixels.  As expected, images obtained in this configuration are
sharper than in the 1.8BF case; nevertheless MEM restoration still
shows significant improvements.  SNR measurements for Group B are
$\mathrm{{SNR}_{orig}}=9.0$ and $\mathrm{{SNR}_{MEM}}=49.8$.  Group
D: $\mathrm{{SNR}_{orig}}=5.4$ and $\mathrm{{SNR}_{MEM}}=18.5$.
Particle group E is not detected at this setting.

\begin{figure}
\begin{center}
\includegraphics[scale=1]{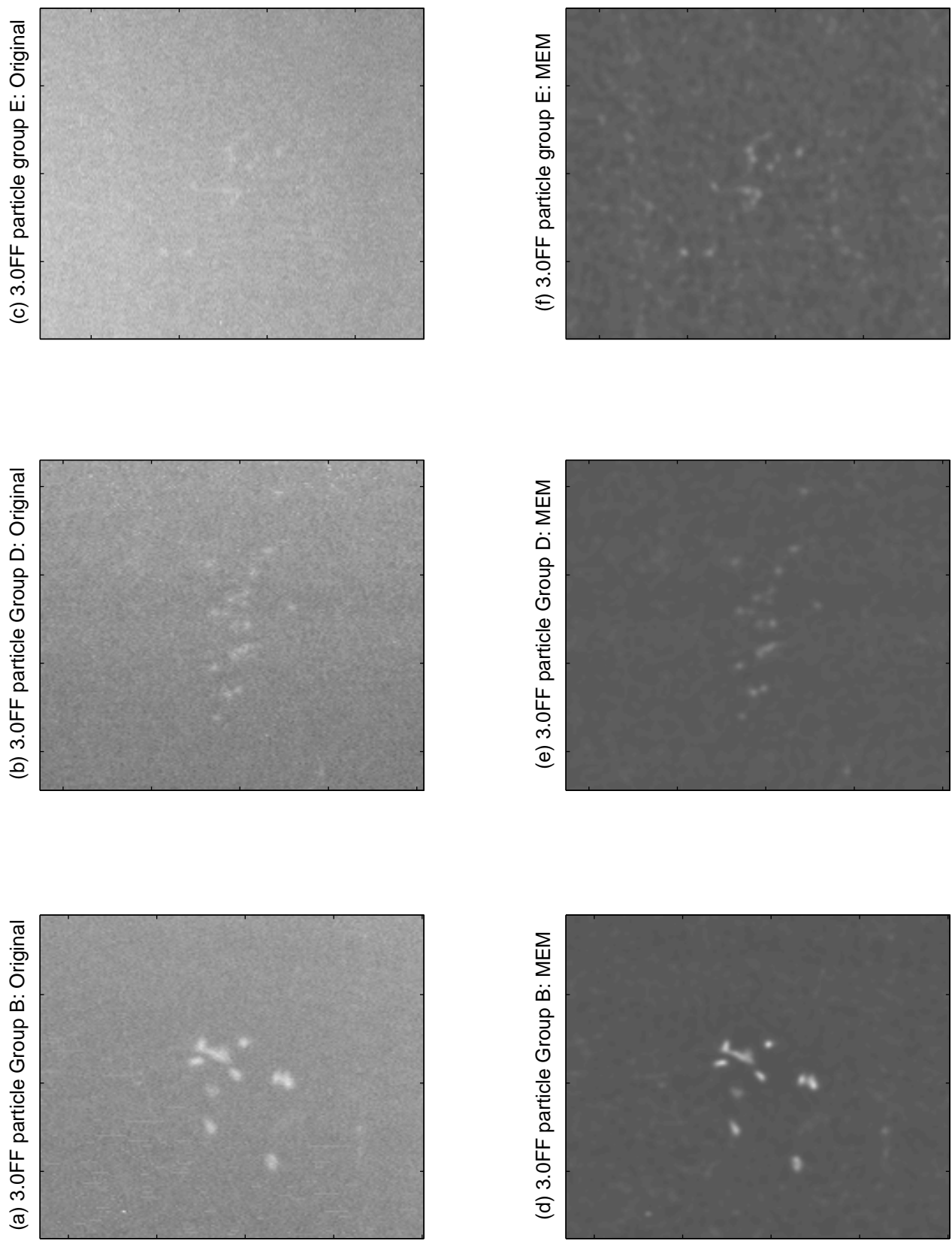}
\caption{3.0FF. Close-ups of particle groups B, D and E are shown
in (a), (b) and (c) respectively.  MEM restorations are shown
beneath in (d), (e) and (f); improvements in visibility and resolution
are apparent in all three cases, with fine details of individual
microcalcifications being discernable after MEM restoration (d).}
\label{figure3.0FF}
\end{center}
\end{figure}

Figure \ref{figure3.0FF} shows cropped images of the same three
particle groups, obtained with an unconventional 3.0FF
configuration. In this case all three groups are detectable in the
original image of the phantom, but the MEM processed images show
clear improvements in resolution, enabling fine details of
individual microcalcifications to be discerned in image (d).  SNR
measurements for Group B are $\mathrm{{SNR}_{orig}}=8.8$ and
$\mathrm{{SNR}_{MEM}}=49.8$.  Group D: $\mathrm{{SNR}_{orig}}=5.8$
and $\mathrm{{SNR}_{MEM}}=22.4$. Group E: $\mathrm{{SNR}_{orig}}=5.2$
and $\mathrm{{SNR}_{MEM}}=10.4$.

To effect the same comparison as that illustrated in figure \ref{figureCALC},
between unprocessed 1.8FF images and MEM processed 1.8BF ones, the bottom row in figure
\ref{figure1.8BF} should be compared with the top row in figure \ref{figure1.8FF}.

\begin{figure}
\begin{center}
\includegraphics[scale=1]{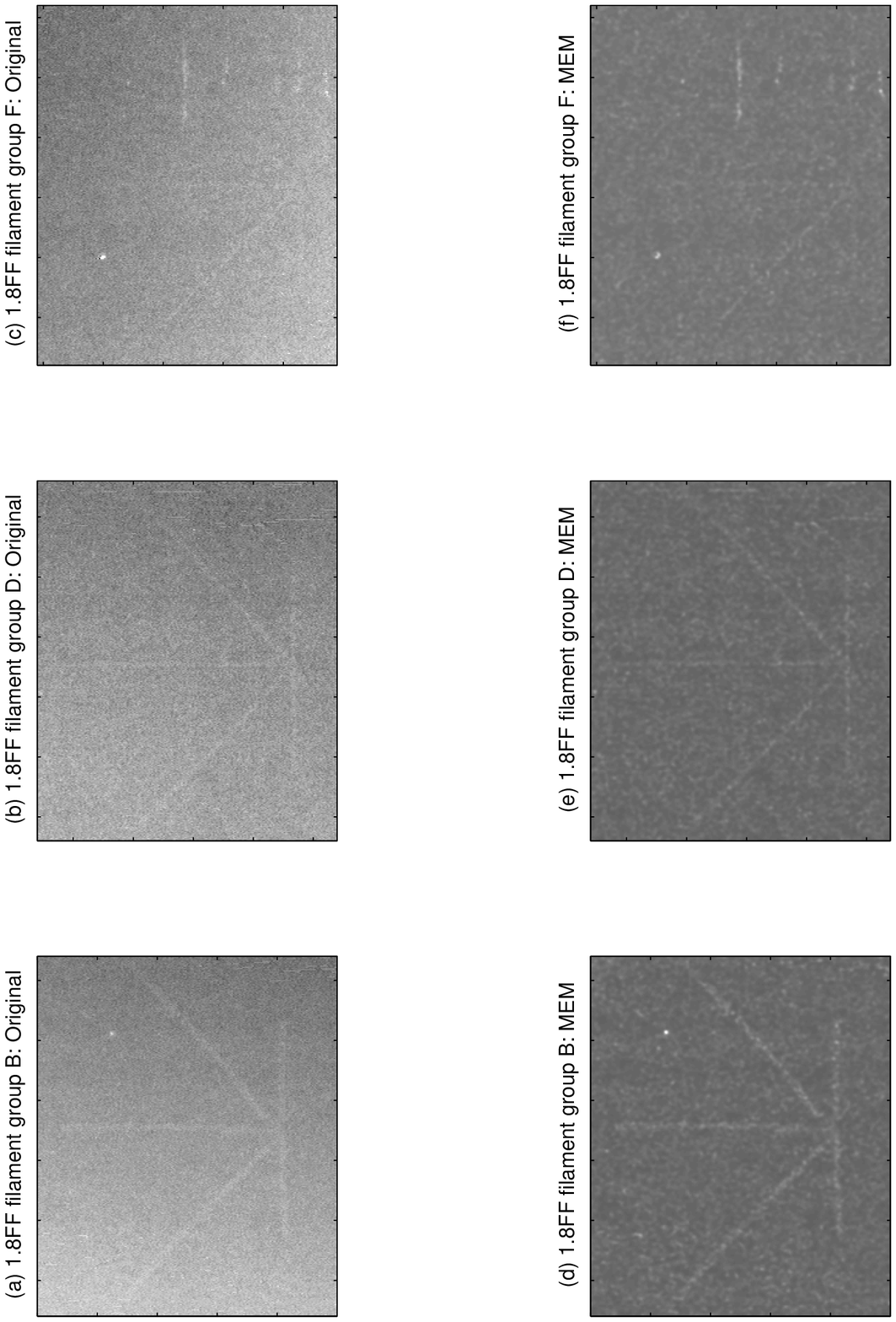}
\caption{Close-ups of filament groups B, D and F are shown
in (a), (b) and (c) respectively.  MEM restorations are shown
beneath in (d), (e) and (f); improvements in visibility and resolution
are apparent in all three cases.}
\label{figureCROW2}
\end{center}
\end{figure}

Figure \ref{figureCROW2} shows images of filament groups B, D and F
taken with the 1.8FF configuration.  These respectively have diameters
0.35, 0.25 and 0.20 mm, length 10 mm.  All three groups are detectable
in the original and MEM processed images, but again the latter
show improvements in resolution and visibility.  To avoid information
overload we do not show the 1.8BF and 3.00FF cases here, but experience
with these matches that with the particle groups, and is quantified
in the table of scores discussed below.

The results obtained from scoring the original and MEM processed
images under each of the three imaging geometries are shown in
Table \ref{Table1}.  The processed images show an improved score
in each case.  This implies that the signal-to-noise ratio
perceived by the observers for the various test features,
including low contrast objects, was increased by the application
of the MEM processing.

\Table{\label{Table1}Image scores of the original and MEM
processed images, obtained by averaging the individual scores of
the two independent observers.} \br
      Mode &     Image  &               \multicolumn{4}{c}{TORMAM scores}  \\
\cline{3-6} \ms
           &            &  Filaments &  Particles &      Disks &      Total \\
\mr
     1.8BF &   Original &      \020.5 &      \0\06.0 &    \024.5 &      \051.0 \\

           &        MEM &      \038.5 &      \0\08.5 &    \030.0 &      \077.0 \\

     1.8FF &   Original &      \027.0 &       \0\06.0 &   \021.5 &      \054.5 \\

           &        MEM &      \041.0 &      \010.5 &     \029.5 &      \081.0 \\

     3.0FF &   Original &      \042.0 &      \0\09.5 &    \027.5 &      \079.0 \\

           &        MEM &      \057.5 &       \011.0 &     \035.5 &       104.0 \\
\br
\endTable

\section {Discussion}

The aim of the work was to demonstrate an improvement in spatial
resolution for realistic radiological images from MEM de-blurring,
with no associated penalty in terms of reduction in the
signal-to-noise ratio perceived by the observer. The expected resolution
improvements are shown in figures \ref{figure1.8BF}-\ref{figureCROW2}.
Somewhat unexpected, however, were the improvements in image score shown in
Table 1, as most of the features in the TORMAM phantom, i.e. the
fibre groups and disks, are essentially low-contrast features
whose detection would be expected to be limited by the relative
noise level in the image. The improvement in scores therefore
implies an improvement in signal-to-noise ratio for this phantom.
For the filaments and particles, which are comparable to the PSF
in extent, improvements in visibility are effected by enhanced
intensity and sharpness due to focussing, and by noise reduction.
For the disks, which are significantly larger than the PSF, sharpness
(resulting in easier edge detection) and noise reduction are the
important factors.

It may be that the improvements in features imaged against
a uniform background, as in these demonstrations, are better than those
which might be achieved when imaging diagnostic features against an
anatomical background.  Suitable experiments are under way.

There are a number of possible applications of MEM deconvolution
in radiology. This paper has demonstrated the use of high
geometric magnification in conjunction with a relatively large
focal spot. Other options might be to use larger focal spot tubes
to increase heat capacity and therefore allow extremely short
exposures, or high outputs which could be used with high
filtration. Other deconvolution problems in radiology include
light scatter in image intensifier optics and reduction of the
effects of scattered radiation.

\section {Conclusion}

An image processing approach based on a maximum entropy method has
been applied to the problem of restoring focal spot geometric
blurring in magnification mammography. The results show an
improvement in image spatial resolution and an improvement in
terms of the image signal-to-noise ratio perceived by the
observer, as evaluated using a standard phantom and at a realistic
quantum noise level. This successful demonstration of image
de-blurring in noisy radiological images offers the possibility of
weakening the link between focal spot size and geometric blurring
in radiology, and thus opening up new approaches to system
optimization.

\section{Acknowledgements}

A. Jannetta acknowledges receipt of a research studentship from
Northumbria University.  We are grateful to The Royal Society for
award of Research Grant 574006.G503/23863/SM for the purchase of
dedicated hardware and image processing software.  It is a pleasure
to thank Dr. Stephen Gull of the Astrophysics Group, Cavendish
Laboratory, University of Cambridge for many useful conversations
about MEM.

The authors acknowledge Leeds Test Objects Ltd for permission to
reproduce a section of the layout of their TORMAM phantom in this
paper.

\References

\harvarditem{Andrews \harvardand\ Hunt}{1977}{ANDREWS} Andrews H~C
\harvardand\ Hunt B~R 1977
\newblock {\em Digital Image Restoration} (Englewood Cliffs, NJ: Prentice-Hall)

\harvarditem{Coleman \harvardand\ Li}{1994}{COLEMAN2}
Coleman T~F \harvardand\ Li Y 1994
\newblock On the convergence of reflective Newton methods for large-scale
nonlinear minimization subject to bounds
{\em Mathematical Programming} {\bf 67}~189--224

\harvarditem{Coleman \harvardand\ Li}{1996}{COLEMAN1}
\dash 1996
\newblock An interior, trust region approach for nonlinear minimization subject to bounds
{\em SIAM J. Optim.} {\bf 6}~418--45

\harvarditem{Cornwell \harvardand\ Evans}{1985}{CORNEVANS}
Cornwell T~J \harvardand\ Evans K~F 1985
\newblock A simple maximum entropy deconvolution algorithm
{\em Astron. Astrophys.} {\bf 143}~77--83

\harvarditem{Cowen {\em et~al}}{1992}{COWEN}
Cowen A~R, Brettle D~S, Coleman N~J \harvardand\ Parkin G~J~S 1992
\newblock A preliminary investigation of the imaging performance of
photostimulable phosphor computed radiography using a new design
of mammographic quality control test object {\em Br. J. Radiol.}
{\bf 65}~528--35

\harvarditem{Csisz\`{a}r}{1991}{CSISZAR} Csisz\`{a}r I 1991
\newblock Why least squares and maximum entropy? An
axiomatic approach to inference for linear inverse problems
{\em The Annals of Statistics} {\bf 19}(4)~2032--66

\harvarditem{Dempster {\em et~al}}{1977}{DEMPSTER}
Dempster A, Laird N \harvardand\ Rubin D 1977
\newblock Maximum likelihood from incomplete data via the {EM} algorithm
{\em J. R. Statist. Soc.} B {\bf 39}(1)~1--38

\harvarditem{De~Pierro}{1995}{DEPIERRO}
De~Pierro A~R 1995
\newblock A modified expectation maximization algorithm for penalized
likelihood estimation in emission tomography {\em IEEE Trans. Med.
Imaging} {\bf 14}(1)~132--7

\harvarditem{Donoho {\em et~al}}{1992}{DONOHO3}
Donoho D~L, Johnstone I~M, Hoch J~C \harvardand\ Stern A~S 1992
\newblock Maximum entropy and the nearly black object
{\em J. R. Statist. Soc.} B {\bf 54}(1)~41--81

\harvarditem{Dougherty \harvardand\ Kawaf}{2001}{DOUGH}
Dougherty G \harvardand\ Kawaf Z 2001
\newblock The point-spread function revisited: image restoration using 2-{D} deconvolution
{\em Radiography} {\bf 7}(4)~255--62

\harvarditem{Frieden}{1972}{FRIEDEN}
Frieden B~R 1972
\newblock Restoring with maximum likelihood and maximum entropy
{\em J. Opt. Soc. Am.} {\bf 62}~511--8

\harvarditem{Gonzalez \harvardand\ Woods}{2001}{GONZALEZ}
Gonzalez R~C \harvardand\ Woods R~E 2001
\newblock {\em Digital Image Processing} (Upper Saddle River, NJ: Prentice-Hall)

\harvarditem{Gull \harvardand\ Daniell}{1978}{GULLDAN}
Gull S~F \harvardand\ Daniell G~J 1978
\newblock Image reconstruction from incomplete and noisy data
{\em Nature} {\bf 272}~686-90

\harvarditem{Gull \harvardand\ Skilling}{1999}{GULLSKILL2}
Gull S~F \harvardand\ Skilling J 1999
\newblock {\em MEMSYS5 Users' Manual} (Bury St. Edmunds:
Maximum Entropy Data Consultants Ltd)

\harvarditem{Helstrom}{1967}{HELSTROM}
Helstrom C~W 1967
\newblock Image restoration by the method of least squares
{\em J. Opt. Soc. Am.} {\bf 57}(3)

\harvarditem{Hudson \harvardand\ Larkin}{1994}{HUDSON}
Hudson H~M \harvardand\ Larkin R~S 1994
\newblock Accelerated image reconstruction using ordered subsets of projection data
{\em IEEE Trans. Med. Imaging} {\bf 13}(4)~601--9

\harvarditem{IRMER}{2000}{IRMER2000} IRMER 2000 The Ionising
Radiation (Medical Exposure) Regulations 2000 (London: HMSO)

\harvarditem{Jackson \harvardand\ Ward}{1981}{JACKSON}
Jackson J~C \harvardand\ Ward G 1981
\newblock Surface inspection of steel products using a synthetic aperture
microwave technique {\em Br. J. Non-Destr. Test.} {\bf
33}(8)~395--402

\harvarditem{Jaynes}{1957{\em a}}{JAYNES4}
Jaynes E~T 1957a \newblock Information theory and statistical mechanics
{\em Phys. Rev.} {\bf 106}~620

\harvarditem{Jaynes}{1957{\em b}}{JAYNES5}
\dash 1957b \newblock Information theory and statistical mechanics {II}
{\em Phys. Rev.} {\bf 108}~171

\harvarditem{Jaynes}{1968}{JAYNES2}
\dash 1968
\newblock Prior probabilities
{\em IEEE Trans. Systems Science and Cybernetics} {\bf 4}(3)~227--41

\harvarditem{Jaynes}{1982}{JAYNES3}
\dash 1982
\newblock On the rationale of maximum entropy methods
{\em Proc. IEEE} {\bf 70}(9)~939--52

\harvarditem{Kinahan {\em et~al}}{1997}{KINAHAN}
Kinahan P~E, Fessler J~A \harvardand\ Karp J~S 1997
\newblock Statistical image reconstruction in PET with compensation
for missing data {\em IEEE Tr. Nucl. Sci.} {\bf 44(4)}~1552--7

\harvarditem{Lucy}{1974}{LUCY}
Lucy L~B 1974
\newblock An iterative technique for the rectification of observed distributions
 {\em Astron. J.} {\bf 79}~745

\harvarditem{O'Sullivan {\em et~al}}{1998}{OSULLIVAN}
O'Sullivan J~A, Blahut R~E \harvardand\ Snyder D~L 1998
\newblock Information-theoretic image formation
{\em IEEE Trans. Information Theory} {\bf 44}(6)~2094--2123

\harvarditem{Richardson}{1972}{RICHARDSON}
Richardson W~H 1972
\newblock Bayesian-based iterative method of image restoration
{\em J. Opt. Soc. Am.} {\bf 62}~55--59

\harvarditem{Shepp \harvardand\ Vardi}{1982}{SHEPP} Shepp L~A
\harvardand\ Vardi Y 1982 \newblock Maximum likelihood
reconstruction for emision tomography {\em IEEE Trans. Med.
Imaging} {\bf 1}(2)~113--22

\harvarditem{Skilling}{1988}{SKILLING}
Skilling J 1988
\newblock The axioms of maximum entropy {\em Maximum Entropy
and Bayesian Methods in Science and Engineering, Volume 1: Foundations}
ed G J Erickson \harvardand\ C~R Smith (Kluwer: Dordrecht) pp~173-87

\harvarditem{Skilling}{1989}{SKILLING2}
\dash 1989
\newblock Classic maximum entropy {\em Maximum Entropy and Bayesian Methods}
ed J Skilling (Kluwer: Dordrecht) p~45

\harvarditem{Skilling \harvardand\ Bryan}{1984}{SKILLINGBRYAN}
Skilling J \harvardand\ Bryan R~K 1984
\newblock Maximum entropy image reconstruction - general algorithm
{\em Mon. Not. R. astr. Soc.} {\bf 211}~111--24

\harvarditem{Skilling \harvardand\ Gull}{1985}{GULLSKILL}
Skilling J \harvardand\ Gull S~F 1985
\newblock Algorithms and applications
{\em Maximum Entropy and Bayesian Methods in Inverse Problems}
ed C~R Smith \harvardand\ W~T Grandy (Reidel: Dordrecht) p~83

\harvarditem{Snyder {\em et~al}}{1992}{SNYDER1}
Snyder D~L, Schulz T~J \harvardand\ O'Sullivan J~A 1992
\newblock Deblurring subject to nonnegativity constraints
{\em IEEE Trans. Signal Processing} {\bf 40}(5)~1143--50

\harvarditem{Wiener}{1949}{WIENER} Wiener N 1949
\newblock {\em Extrapolation, interpolation and smoothing of stationary time
series} (Cambridge: MIT Press and New York: Wiley)

\harvarditem{Young \harvardand\ Ramsdale}{2003}{NHSBSP2003}
Young K~C \harvardand\ Ramsdale M~L 2003
\newblock {\em Performance of mammographic equipment in the UK Breast Screening Programme in 2000/2001} (NHS Breast Screening Programme Publication 56)

\endrefs

\end{document}